\documentclass[12pt,thmsa]{article}

\def \eq {\begin{equation}}
\def \fim-eq {\end{equation}}
\begin{document}

\author{E. S. Guerra \\
Departamento de F\'{\i}sica \\
Universidade Federal Rural do Rio de Janeiro \\
Cx. Postal 23851, 23890-000 Serop\'edica, RJ, Brazil \\
email: emerson@ufrrj.br\\
}
\title{ TELEPORTATION OF ATOMIC STATES FOR ATOMS IN A LAMBDA CONFIGURATION}
\maketitle

\begin{abstract}
\noindent In this article we discuss a scheme of teleportation of atomic
states making use of three-level lambda atoms. The experimental realization
proposed makes use of cavity QED involving the interaction of Rydberg atoms
with a micromaser cavity prepared in a coherent state. We start presenting a
scheme to prepare atomic EPR states involving two-level atoms via the
interaction of these atoms with a cavity. In our scheme the cavity and some
atoms play the role of auxiliary systems used \ to achieve the teleportation.

\ \newline

\end{abstract}

\section{\protect\bigskip INTRODUCTION}

Both entanglement and non-locality, which are closely related, are important
concepts in quantum mechanics with applications in information processing
and quantum computing \cite{Nielsen, MathQC}. One of the most dramatic among
various consequences of entanglement and non-locality with applications in
information science is teleportation, put forward by Bennett \textit{at al} 
\cite{Bennett}. The essentials of the teleportation scheme is that, given an
unknown quantum state to the sender, making use of quantum entanglement and
non-locality, it is possible \ to reproduce this state far apart in the
quantum system of the receiver where, in the process, both the sender and
the receiver follow a certain prescription and communicate with each other
through a classical channel. In the end of the process the receiver has a
quantum state similar to the quantum state of the sender and the quantum
state of the sender is destroyed since, according to the no-cloning theorem 
\cite{Nocloning, Nielsen} it is not possible to clone a quantum state.
Quantum teleportation is an experimental reality and it holds tremendous
potential for applications in the fields of quantum communication and
computing \cite{MathQC, Nielsen}. For instance, it can be used \ to build
quantum gates which are resistant to noise and is intimately connected with
quantum error-correcting codes \cite{Nielsen}. The most significant
difficulty for quantum teleportation to become an useful tool in quantum
communication and computing is how to avoid decoherence effects \cite%
{Orszag, Zurek}. A scheme of teleportation of atomic states, using cavity
QED, has been proposed in Ref.\cite{david}. For several proposals of
realization schemes of teleportation see \cite{MathQC}.

In this article we present a scheme of teleportation close to the original
scheme presented by Bennett \textit{et at } \cite{Bennett}. We will assume
that Alice and Bob meet and then create an EPR atomic state \ involving
atoms $A2$ and $\ A4$. Then Alice and Bob separate. Alice takes with her
half of the EPR pair, that is, atom $A2$ and Bob keeps with him the other
half of the EPR pair, that is, atom $A4$. Later on Alice are going to be
able to teleport to Bob's atom $A4$ an unknown state of an atom $A1$ making
use of her half of the EPR pair, that is, atom $A2$.

In the discussion which follows we are going to consider Rydberg atoms of
relatively long radiative lifetimes \cite{Rydat}. We also assume a perfect \
microwave cavity and we neglect effects due to decoherence. Concerning this
point, it is worth to mention that nowadays it is possible to build up
niobium superconducting cavities with high quality factors $Q$. It is
possible to construct cavities with quality factors $Q\sim 10^{8}$ \cite%
{haroche}. Even cavities with quality factors as high as $Q\sim 10^{12}$
have been reported \cite{walther}, which, for frequencies $\nu \sim 50$ GHz
gives us a cavity field lifetime of the order of a few seconds.

\section{EPR STATES}

Let us first show how we can get an EPR state \cite{EPR} making use of
three-level lambda atoms interacting with a cavity field. Consider a
three-level lambda atom (see Fig. 1) interacting with the electromagnetic
field inside a cavity $C$. The states of \ the atom, $|a\rangle ,$ $%
|b\rangle $ and $|c\rangle $ are so that the $|a\rangle \rightleftharpoons
|c\rangle $ and $|a\rangle \rightleftharpoons |b\rangle $ transitions are in
the far off resonance interaction limit. The time evolution operator for the
atom-field interaction $U(t)$ is given by \cite{Knight} 
\begin{equation}
U(\tau )=-e^{i\varphi a^{\dagger }a}|a\rangle \langle a|+\frac{1}{2}%
(e^{i\varphi a^{\dagger }a}+1)|b\rangle \langle b|+\frac{1}{2}(e^{i\varphi
a^{\dagger }a}-1)|b\rangle \langle c|\ +\frac{1}{2}(e^{i\varphi a^{\dagger
}a}-1)|c\rangle \langle b|+\frac{1}{2}(e^{i\varphi a^{\dagger
}a}+1)|c\rangle \langle c|,  \label{U1lambda}
\end{equation}%
where $a$ $(a^{\dagger })$ is the annihilation (creation) operator for the
field in cavity $C$, $\varphi =2g^{2}\tau /$ $\Delta $, \ $g$ is the
coupling constant, $\Delta =\omega _{a}-\omega _{b}-\omega =\omega
_{a}-\omega _{c}-\omega $ is the detuning where \ $\omega _{a}$, $\omega
_{b} $ and $\omega _{c}$\ are the frequency of the upper level and \ of the
two degenerate lower levels respectively and $\omega $ is the cavity field
frequancy and $\tau $ is the atom-field interaction time. For $\varphi =\pi $%
, we get 
\begin{equation}
U(\tau )=-\exp \left( i\pi a^{\dagger }a\right) |a\rangle \langle a|+\Pi
_{+}|b\rangle \langle b|+\Pi _{-}|b\rangle \langle c|\ +\Pi _{-}|c\rangle
\langle b|+\Pi _{+}|c\rangle \langle c|,  \label{UlambdaPi}
\end{equation}%
where 
\begin{eqnarray}
\Pi _{+} &=&\frac{1}{2}(e^{i\pi a^{\dagger }a}+1),  \nonumber \\
\Pi _{-} &=&\frac{1}{2}(e^{i\pi a^{\dagger }a}-1).  \label{pi+-}
\end{eqnarray}

Considering the non-normalized even and odd coherent states \cite{EvenOddCS} 
\begin{equation}
|\pm \rangle =|\alpha \rangle \pm |-\alpha \rangle ,  \label{+-}
\end{equation}%
in the following calculations we shall use the relations 
\begin{eqnarray}
\Pi _{+}|+\rangle &=&|+\rangle ,  \nonumber \\
\Pi _{+}|-\rangle &=&0,  \nonumber \\
\Pi _{-}|-\rangle &=&-|-\rangle ,  \nonumber \\
\Pi _{-}|+\rangle &=&0,
\end{eqnarray}%
which are easily obtained from Eqs. (\ref{pi+-}) and (\ref{+-}), using $%
e^{za^{\dagger }a}|\alpha \rangle =|e^{z}\alpha \rangle $ \cite{Louisell}.

Let us prepare the cavity $C$ in the coherent state $|\alpha \rangle $ and
consider the atom $A1$ in the following state 
\[
|\psi \rangle _{A1}=|b_{1}\rangle . 
\]%
The initial state of the atom-cavity system is given by%
\begin{equation}
|\psi (0)\rangle _{A1-C}=|\psi \rangle _{A1}|\alpha \rangle =|b_{1}\rangle
|\alpha \rangle =|b_{1}\rangle \frac{1}{2}[|+\rangle +|-\rangle ].
\end{equation}%
We now let atom $A1$ fly through the cavity $C$. The state \ of the system
evolves according to the time evolution operator Eq. (\ref{UlambdaPi})
yielding 
\begin{eqnarray}
|\psi (\tau )\rangle _{A1-C} &=&U(\tau )|\psi (0)\rangle _{A0-C}  \nonumber
\\
&=&\frac{1}{2}\{|b_{1}\rangle |+\rangle -|c_{1}\rangle |-\rangle \}.
\end{eqnarray}

Consider now another three-level lambda atom $A2$ prepared initially in the
state $|b_{2}\rangle $, which are going to pass through the cavity. Now, as
initial state of the system, we have 
\begin{equation}
|\psi (0)\rangle _{A1-A2-C}=\frac{1}{2}\{|b_{1}\rangle |+\rangle
-|c_{1}\rangle |-\rangle \}|b_{2}\rangle .
\end{equation}%
After this second atom has passed through the cavity, the system evolves to 
\begin{eqnarray}
|\psi (\tau )\rangle _{A1-A2-C} &=&U(\tau )|\psi (0)\rangle _{A1-C-A2} 
\nonumber \\
&=&\frac{1}{2}\{|b_{1}\rangle |b_{2}\rangle |+\rangle +|c_{1}\rangle
|c_{2}\rangle |-\rangle \}.  \label{LBPSIA1A2C}
\end{eqnarray}%
Now, we inject a coherent state $|\alpha \rangle $ in the cavity, that is,
we make use of $D(\beta )|\alpha \rangle =|\alpha +\beta \rangle $, and we
get 
\begin{eqnarray}
|\psi \rangle _{A1-A2-C} &=&\frac{1}{2}\{|b_{1}\rangle |b_{2}\rangle
(|2\alpha \rangle +|0\rangle )+|c_{1}\rangle |c_{2}\rangle (|2\alpha \rangle
-|0\rangle )\}  \nonumber \\
&=&\frac{1}{2}\{(|b_{1}\rangle |b_{2}\rangle +|c_{1}\rangle |c_{2}\rangle
)|2\alpha \rangle +(|b_{1}\rangle |b_{2}\rangle -|c_{1}\rangle |c_{2}\rangle
)|0\rangle \}.
\end{eqnarray}%
In order to disentangle the atomic states of the cavity field state we now
send a two-level atom $A3,$ resonant with the cavity, with $|f_{3}\rangle $
and $|e_{3}\rangle $ being the lower and upper levels respectively, through $%
C$. If $A3$ is sent in the lower state $|f_{3}\rangle $, under the
Jaynes-Cummings dynamics \cite{Orszag} we know that the state $|f_{3}\rangle
|0\rangle $ does not evolve, however, the state $|f_{3}\rangle |2\alpha
\rangle $ evolves to $|e_{3}\rangle |\chi _{e}\rangle +|f_{3}\rangle |\chi
_{f}\rangle $, where $|\chi _{f}\rangle =\sum\limits_{n}C_{n}\cos (gt\sqrt{n}%
)|n\rangle $ and $|\chi _{e}\rangle =-i\sum\limits_{n}C_{n+1}\sin (gt\sqrt{%
n+1})|n\rangle $ and $C_{n}=e^{-\frac{1}{2}|2\alpha |^{2}}(-2\alpha )^{n}/%
\sqrt{n!}$. Then we get%
\begin{equation}
\mid \psi \rangle _{A1-A2-A3-C}=\frac{1}{2}\{(|b_{1}\rangle |b_{2}\rangle
+|c_{1}\rangle |c_{2}\rangle )(|e_{3}\rangle |\chi _{e}\rangle
+|f_{3}\rangle |\chi _{f}\rangle )+(|b_{1}\rangle |b_{2}\rangle
-|c_{1}\rangle |c_{2}\rangle )|f_{3}\rangle |0\rangle \},
\end{equation}%
and if we detect atom $A3$ in state $|e_{3}\rangle ,$ then we finally get
the EPR state involving the atoms $A1$ and $A2$ 
\begin{equation}
\mid \Psi ^{+}\rangle _{A1-A2}=\frac{1}{\sqrt{2}}(|b_{1}\rangle
|b_{2}\rangle +|c_{1}\rangle |c_{2}\rangle ),  \label{LBBellPSI+}
\end{equation}%
which is an entangled state of atoms $A1$ and $A2$.

In the above disentanglement process we can choose a coherent field with a
photon-number distribution with a sharp peak at average photon number $%
\langle n\rangle =|\alpha |^{2}$ so that, to a good approximation, $|\chi
_{f}\rangle \cong C_{\overline{n}}\cos (\sqrt{\overline{n}}g\tau )|\overline{%
n}\rangle $ and $|\chi _{e}\rangle \cong C_{\overline{n}}\sin (\sqrt{%
\overline{n}}g\tau )|\overline{n}\rangle $, where $\overline{n}$ is the
integer nearest $\langle n\rangle $, and we could choose, for instance $\ 
\sqrt{\overline{n}}g\tau =\pi /2$, so that we would have $|\chi _{e}\rangle
\cong C_{\overline{n}}|\overline{n}\rangle $ and $|\chi _{f}\rangle \cong 0$%
. In this case we atom $A3$ \ would be detected in state $|e_{3}\rangle $
with almost $100\%$ of probability. Therefore, proceeding this way, we can
guarantee that the atomic and field states will be disentangled successfully
as we would like.

Now, if we start from (\ref{LBPSIA1A2C}) and we inject a coherent state $%
|-\alpha \rangle $ in the cavity and let a two-level atom $A3$ resonant with
the cavity to fly through the cavity prepared initially in the state lower $%
|f_{3}\rangle $ and detect the upper state $|e_{3}\rangle ,$ we get%
\begin{equation}
\mid \Psi ^{-}\rangle _{A1-A2}=\frac{1}{\sqrt{2}}(|b_{1}\rangle
|b_{2}\rangle -|c_{1}\rangle |c_{2}\rangle ),  \label{LBBellPSI-}
\end{equation}%
which is also an EPR state.

Now, if we apply the rotation%
\begin{equation}
R=\left[ 
\begin{array}{cc}
0 & -1 \\ 
1 & 0%
\end{array}%
\right] ,  \label{R1}
\end{equation}%
that is,%
\begin{equation}
R=\mid c_{2}\rangle \langle b_{2}|-\mid b_{2}\rangle \langle c_{2}|,
\label{R2}
\end{equation}%
to the state (\ref{LBBellPSI+}) we get 
\begin{equation}
\mid \Phi ^{-}\rangle _{A1-A2}=\frac{1}{\sqrt{2}}(|b_{1}\rangle
|c_{2}\rangle -|c_{1}\rangle |b_{2}\rangle ),  \label{LBBellPHI-}
\end{equation}%
and applying (\ref{R2}) to the state (\ref{LBBellPSI-}) we get%
\begin{equation}
\mid \Phi ^{+}\rangle _{A1-A2}=\frac{1}{\sqrt{2}}(|b_{1}\rangle
|c_{2}\rangle +|c_{1}\rangle |b_{2}\rangle ).  \label{LBBellPHI+}
\end{equation}%
The rotation of the atomic states (\ref{R1}) can be performed as shown in 
\cite{GHZLambdaat}. The states (\ref{LBBellPSI+}), (\ref{LBBellPSI-}), (\ref%
{LBBellPHI-}) and (\ref{LBBellPHI+}) form a Bell basis \cite{BELLbasis}.

\section{TELEPORTATION}

In this section we are going discuss a teleportation scheme that is closely
similar to the original scheme suggested by Bennett \textit{et al} \cite%
{Bennett}. Let us assume that Alice and Bob meet and than they build up an
EPR state involving two-level atoms $A2$ and $A4$ as described in section 2
(we use the notation $A3$ for the two-level atom used to disentangle the
atomic states from the cavity state as in the previous sections). That is,
as in section 2 they make use of a cavity prepared initially in a coherent
state $\mid \alpha \rangle $ and by sending $A2$ and $A4$ through this
cavity where the atoms interact dispersively with the cavity, and following
the recipe presented in that section they get

\begin{equation}
\mid \Psi ^{+}\rangle _{A2-A4}=\frac{1}{\sqrt{2}}(\mid b_{2}\rangle \mid
b_{4}\rangle +\mid c_{2}\rangle \mid c_{4}\rangle ),
\end{equation}%
Now, let us assume that Alice keeps with her the half of this EPR state
consisting of atom $A2$ and Bob keeps with him the other half of this EPR
state, that is, atom $A4$. Then they separate and let us assume that they
are far apart from each other. Later on, Alice decides to teleport the state
of an atom $A1$ prepared in an unknown state%
\begin{equation}
\mid \psi \rangle _{A1}=\zeta \mid b_{1}\rangle +\xi \mid c_{1}\rangle ,
\end{equation}%
to Bob. The initial state of the system $A1-A2-A4$ is%
\begin{eqnarray}
&\mid &\psi \rangle _{A1-A2-A4}=\frac{1}{\sqrt{2}}\{\zeta (\mid b_{1}\rangle
\mid b_{2}\rangle \mid b_{4}\rangle +\mid b_{1}\rangle \mid c_{2}\rangle
\mid c_{4}\rangle )+  \nonumber \\
\xi ( &\mid &c_{1}\rangle \mid b_{2}\rangle \mid b_{4}\rangle +\mid
c_{1}\rangle \mid c_{2}\rangle \mid c_{4}\rangle )\}.  \label{Initstate}
\end{eqnarray}%
Then, first Alice prepares a cavity $C$ in a coherent state $\mid \alpha
\rangle $ and sends $A1$ and $A2$ through $C$. Taking into account (\ref%
{UlambdaPi}) with $\varphi =\pi ,$ after atoms $A1$ and $A2$ fly through the
cavity she gets%
\begin{eqnarray}
&\mid &\psi \rangle _{A1-A2-A4-C}=  \nonumber \\
\frac{1}{\sqrt{2}}\{\zeta \lbrack ( &\mid &b_{1}\rangle \mid b_{2}\rangle
\mid +\rangle +\mid c_{1}\rangle \mid c_{2}\rangle \mid -\rangle )\mid
b_{4}\rangle +  \nonumber \\
( &\mid &b_{1}\rangle \mid c_{2}\rangle \mid +\rangle +\mid c_{1}\rangle
\mid b_{2}\rangle \mid -\rangle )\mid c_{4}\rangle )]+  \nonumber \\
\xi \lbrack ( &\mid &c_{1}\rangle \mid b_{2}\rangle \mid +\rangle +\mid
b_{1}\rangle \mid c_{2}\rangle \mid -\rangle )\mid b_{4}\rangle +  \nonumber
\\
( &\mid &c_{1}\rangle \mid c_{2}\rangle \mid +\rangle +\mid b_{1}\rangle
\mid b_{2}\rangle \mid -\rangle )\mid c_{4}\rangle )]\},
\end{eqnarray}%
or%
\begin{eqnarray}
&\mid &\psi \rangle _{A1-A2-A4-C}=  \nonumber \\
\frac{1}{\sqrt{2}}\{ &\mid &b_{1}\rangle \mid b_{2}\rangle \lbrack \zeta
\mid +\rangle \mid b_{4}\rangle +\xi \mid -\rangle \mid c_{4}\rangle ]+ 
\nonumber \\
&\mid &c_{1}\rangle \mid c_{2}\rangle \lbrack \xi \mid +\rangle \mid
c_{4}\rangle +\zeta \mid -\rangle \mid b_{4}\rangle ]+  \nonumber \\
&\mid &b_{1}\rangle \mid c_{2}\rangle \lbrack \zeta \mid +\rangle \mid
c_{4}\rangle +\xi \mid -\rangle \mid b_{4}\rangle ]+  \nonumber \\
&\mid &c_{1}\rangle \mid b_{2}\rangle \lbrack \xi \mid +\rangle \mid
b_{4}\rangle +\zeta \mid -\rangle \mid c_{4}\rangle ]\}.  \label{PSIA1A2A4C}
\end{eqnarray}%
Now, all Alice has to do is to inject $\mid \alpha \rangle $ \ in the
cavity, send a two-level atom $A3$ resonant with the cavity in the lower
state $\mid f_{3}\rangle $ through the cavity and detect the upper state $%
\mid e_{3}\rangle $ and detect $(\mid b_{1}\rangle \mid b_{2}\rangle ),(\mid
c_{1}\rangle \mid c_{2}\rangle ),(\mid b_{1}\rangle \mid c_{2}\rangle )$ or $%
(\mid c_{1}\rangle \mid b_{2}\rangle )$ and call Bob informing him the
result of her atomic detection. If she detects $(\mid b_{1}\rangle \mid
b_{2}\rangle )$%
\begin{equation}
\mid \psi \rangle _{A4}=\zeta \mid b_{4}\rangle +\xi \mid c_{4}\rangle ,
\end{equation}%
and Bob has got the right state and has to do nothing else.

If Alice detects $(\mid c_{1}\rangle \mid c_{2}\rangle )$ Bob gets%
\begin{equation}
\mid \psi \rangle _{A4}=\zeta \mid b_{4}\rangle +\xi \mid c_{4}\rangle ,
\end{equation}%
and again he has got the right state.

If Alice detects $(\mid b_{1}\rangle \mid c_{2}\rangle )$ Bob gets%
\begin{equation}
\mid \psi \rangle _{A4}=\xi \mid b_{4}\rangle +\zeta \mid c_{4}\rangle ,
\end{equation}%
and he needs to perform the rotation%
\begin{equation}
R_{4}=\left[ 
\begin{array}{cc}
0 & 1 \\ 
1 & 0%
\end{array}%
\right] ,  \label{RotR}
\end{equation}%
on the states of $A4$ and he gets the right state. The rotation (\ref{RotR})
can be performed with an intense field for which $a\longrightarrow \sqrt{n}%
e^{i\theta },a^{\dagger }\longrightarrow \sqrt{n}e^{-i\theta },$ and $%
\varphi a^{\dagger }a\longrightarrow \varphi n$ with $\varphi n=\pi $ in (%
\ref{U1lambda}).

Finally if Alice detects $(\mid c_{1}\rangle \mid b_{2}\rangle )$ Bob gets%
\begin{equation}
\mid \psi \rangle _{A4}=\xi \mid b_{4}\rangle +\zeta \mid c_{4}\rangle ,
\end{equation}%
and again he needs to perform the rotation (\ref{RotR}) on the states of $A4$
and he gets the right state. In Fig. 2 we show the set-up of the
teleportation scheme discussed here.

We can also present a representation \ in terms of qbits of the
teleportation we have

discussed. Making use of%
\begin{eqnarray}
&\mid &b_{k}\rangle =\mid 0_{k}\rangle _{A},  \nonumber \\
&\mid &c_{k}\rangle )=\mid 1_{k}\rangle _{A},  \nonumber \\
&\mid &+\rangle =\mid 0\rangle _{C},  \nonumber \\
&\mid &-\rangle =\mid 1\rangle _{C},
\end{eqnarray}%
we have%
\begin{eqnarray}
|\alpha \rangle &=&\frac{1}{2}(|+\rangle +|-\rangle )=\frac{1}{2}(\mid
0\rangle _{C}+\mid 1\rangle _{C}),  \nonumber \\
|-\alpha \rangle &=&\frac{1}{2}(|+\rangle -|-\rangle )=\frac{1}{2}(\mid
0\rangle _{C}-\mid 1\rangle _{C}),
\end{eqnarray}%
we can write for the initial state of our system taking into account (\ref%
{Initstate})) and the fact that the cavity is initially in a coheret state $%
|\alpha \rangle $ as%
\begin{eqnarray}
&\mid &\psi \rangle _{A1-A2-A4-C}=\frac{1}{2}\{\zeta (\mid 0_{1}\rangle \mid
0_{2}\rangle \mid 0_{4}\rangle +\mid 0_{1}\rangle \mid 1_{2}\rangle \mid
1_{4}\rangle )+  \nonumber \\
\xi ( &\mid &1_{1}\rangle \mid 0_{2}\rangle \mid 0_{4}\rangle +\mid
1_{1}\rangle \mid 1_{2}\rangle \mid 1_{4}\rangle )(\mid 0\rangle _{C}+\mid
1\rangle _{C}).
\end{eqnarray}%
The atom-field interaction in terms of qbits is represented as 
\begin{equation}
\mid A\rangle \mid B\rangle \mid C\rangle \longrightarrow \mid A\oplus
C\rangle \mid B\oplus C\rangle \mid C\rangle ,
\end{equation}%
where the symbol $\oplus $ means addition modulo 2 or it refers to the
logical exclusive-OR (XOR) operation, which is a three qbit quantum gate and
(\ref{PSIA1A2A4C}) can be written as%
\begin{eqnarray}
&\mid &\psi \rangle _{A1-A2-A4-C}=  \nonumber \\
\frac{1}{\sqrt{2}}\{ &\mid &0_{1}\rangle _{A}\mid 0_{2}\rangle _{A}[\zeta
\mid 0\rangle _{C}\mid 0_{4}\rangle _{A}+\xi \mid 1\rangle _{C}\mid
1_{4}\rangle _{A}]+  \nonumber \\
&\mid &1_{1}\rangle _{A}\mid 1_{2}\rangle _{A}[\xi \mid 0\rangle _{C}\mid
1_{4}\rangle _{A}+\zeta \mid 1\rangle _{C}\mid 0_{4}\rangle _{A}]+  \nonumber
\\
&\mid &0_{1}\rangle _{A}\mid 1_{2}\rangle _{A}[\zeta \mid 0\rangle _{C}\mid
1_{4}\rangle _{A}+\xi \mid 1\rangle _{C}\mid 0_{4}\rangle _{A}]+  \nonumber
\\
&\mid &1_{1}\rangle _{A}\mid 0_{2}\rangle _{A}[\xi \mid 0\rangle _{C}\mid
0_{4}\rangle _{A}+\zeta \mid 1\rangle _{C}\mid 1_{4}\rangle _{A}]\}.
\end{eqnarray}%
Then the teleportation process is accomplished detecting $(\mid 0_{1}\rangle
_{A}\mid 0_{2}\rangle _{A}),(\mid 1_{1}\rangle _{A}\mid 1_{2}\rangle
_{A}),(\mid 0_{1}\rangle _{A}\mid 1_{2}\rangle _{A})$ or $(\mid 1_{1}\rangle
_{A}\mid 0_{2}\rangle _{A}$ followed by a rotation of the atomic states if
necessary as described above.

\bigskip \textbf{Figure Captions} \newline

\textbf{Fig. 1-} Energy level scheme of the three-level lambda atom where $%
|a\rangle $ is the upper state with atomic frequency $\omega _{a}$, $%
|b\rangle $ \ and $|c\rangle $ are the lower states with atomic frequency $%
\omega _{b}$ and $\omega _{c}$, $\omega $ is the cavity field frequency and $%
\Delta =\omega _{a}-\omega _{b}-\omega =\omega _{a}-\omega _{c}-\omega $ is
the detuning.\newline

\textbf{Fig. 2- }Alice and Bob meet and generate an EPR state involving
atoms $A2$ and $A4$. Then they separate and Alice keeps atom $A2$ with her
and Bob keeps atom $A4$ with him. Later on Alice decides to teleport an
unknown state prepared in atom $A1$ to Bob. She sends atoms $A1$ and $A2$
through a cavity $C$ prepared initially in a coherent state $|\alpha \rangle 
$. After atoms $A1$ and $A2$ have flown through $C$ Alice injects $|\alpha
\rangle ,$ sends a two-level atom $A3$ prepared initially in the lower state 
$\mid f_{3}\rangle $ \ and resonant with the cavity through $C$ and detects
the upper state $\mid e_{3}\rangle .$ Then she detects the states of atoms $%
A1$ and $A2$ and calls Bob and inform him the result of her atomic
detections. Depending on the results of Alice's atomic detections Bob has or
not to perform an extra rotation $R_{4}$ on the states of his atom $A4.$%
\bigskip \newline

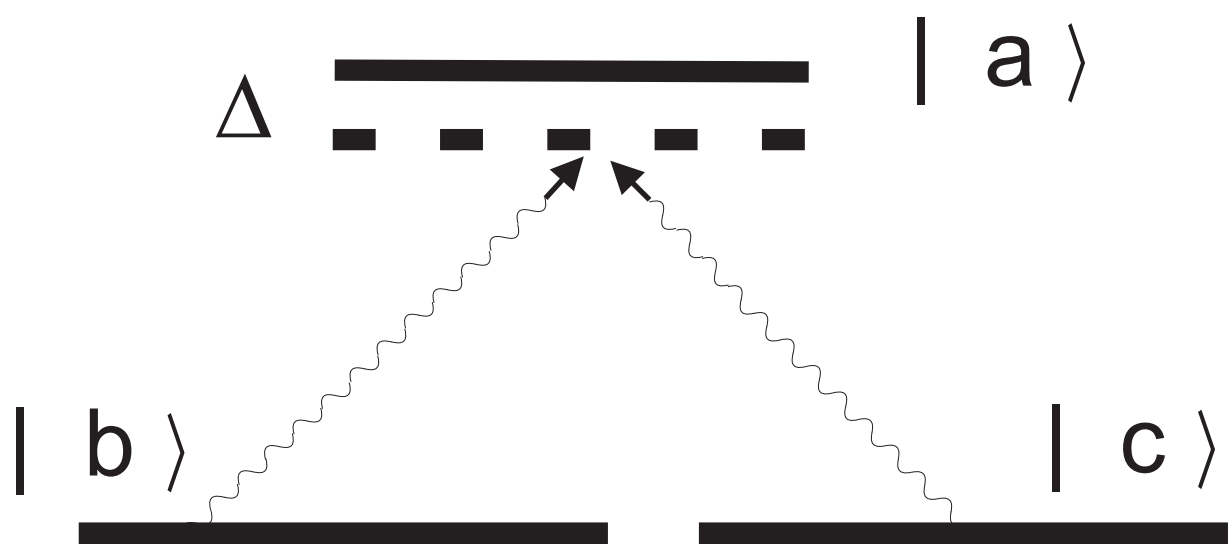
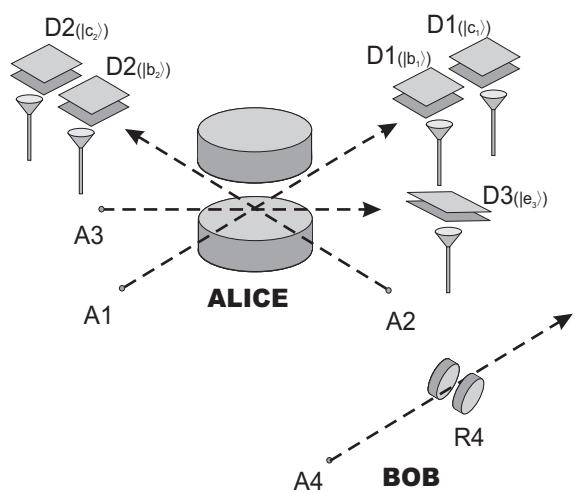
\end{document}